# Adaptive Trajectory Estimation with Power Limited Steering Model under Perturbation Compensation

Weipeng Li, Xiaogang Yang, Ruitao Lu, Jiwei Fan, Tao Zhang, and Chuan He

***Abstract*—Trajectory estimation of maneuvering objects is applied in numerous tasks like navigation, path planning and visual tracking. Many previous works get impressive results in the strictly controlled condition with accurate prior statistics and dedicated dynamic model for certain object. But in challenging conditions without dedicated dynamic model and precise prior statistics, the performance of these methods significantly declines. To solve the problem, a dynamic model called the power-limited steering model (PLS) is proposed to describe the motion of non-cooperative object. It is a natural combination of instantaneous power and instantaneous angular velocity, which relies on the nonlinearity instead of the state switching probability to achieve switching of states. And the renormalization group is introduced to compensate the nonlinear effect of perturbation in PLS model. For robust and efficient trajectory estimation, an adaptive trajectory estimation (AdaTE) algorithm is proposed. By updating the statistics and truncation time online, it corrects the estimation error caused by biased prior statistics and observation drift, while reducing the computational complexity lower than O(n). The experiment of trajectory estimation demonstrates the convergence of AdaTE, and the better robust to the biased prior statistics and the observation drift compared with EKF, UKF and sparse MAP. Other experiments demonstrate through slight modification, AdaTE can also be applied to local navigation in random obstacle environment, and trajectory optimization in visual tracking.**

*Index Terms*—adaptive trajectory estimation, biased prior statistics, dynamic model, perturbation compensation

## I. Introduction

Trajectory estimation plays a critical role from industrial appliances to research areas. It is widely used in numerous tasks like path planning [1], navigation [2], visual tracking [3] and simultaneous localization and mapping (SLAM) [4]-[6]. Successful trajectory estimation depends on three key aspects: dynamic model, measurement model, and estimation algorithm.

Many previous works [7]-[10] have got impressive results under the strictly controlled conditions, where dedicated dynamic model for certain system [11][12], a precise measurement model, and accurate prior statistics can be obtained. However, in some situations where dedicated dynamic model is unobtainable and prior statistics are inaccurate, these methods may be infeasible. Meanwhile, some sensors may experience observation drift. Without any correction, it would lead to serious biased estimation. In this paper, we try to solve the problems above and embark upon two aspects: the design of dynamic model, and robust estimation algorithm.

To describe the movement of non-cooperative or maneuvering objects for the tasks with randomness such as radar tracking and pedestrian path prediction, many outstanding works have created various general dynamic models for maneuvering object tracking. A large category of them is the random model, which is a combination of state evolution model and random control. It contains [13]: wiener-process acceleration model [2], Markov process models such as Singer model [14], and semi-Markov jump process models [15]. These models try to describe the typical states and their switching during maneuver, and get excellent predictions when all of the quantized levels and corresponding probabilities are well designed. But in sever conditions where dedicated dynamic model is unobtainable and prior statistics are inaccurate, they will face two problems. (a) Some models urgently rely on the switching probability between quantized levels, which is a stringent condition in an environment lacking reliable prior information. (b) In most of these models, the movements in orthogonal directions are assumed to be uncoupled with each other, which will weaken the ability of trajectory prediction in many cases such as turning, and split-s maneuver.

Aiming at sever conditions where dedicated dynamic model is unobtainable and prior statistics are inaccurate, we learn from the previous works and propose the power-limited steering model (PLS). It is a natural combination of instantaneous power theory and instantaneous angular velocity theory of Newtonian mechanics. Through the joint action of power and damping, it overcomes the infinite speed problem that occurred in constant mean acceleration model [16] in trajectory prediction. Resort to the strong nonlinearity, PLS needs fewer parameters to describe the switch of typical states compared with Markov jump models.

W. Li, X. Yang, J. Fang, T. Zhang and C. He are with the Department of Control, High-tech Institute of Xi'an, China.
R. Lu. is with the Department of Control, High-tech Institute of Xi'an, Xi'an 710025, and Shaanxi Key Laboratory of Integrated and Intelligent Navigation, China (e-mail: lrt19880220@163.com)

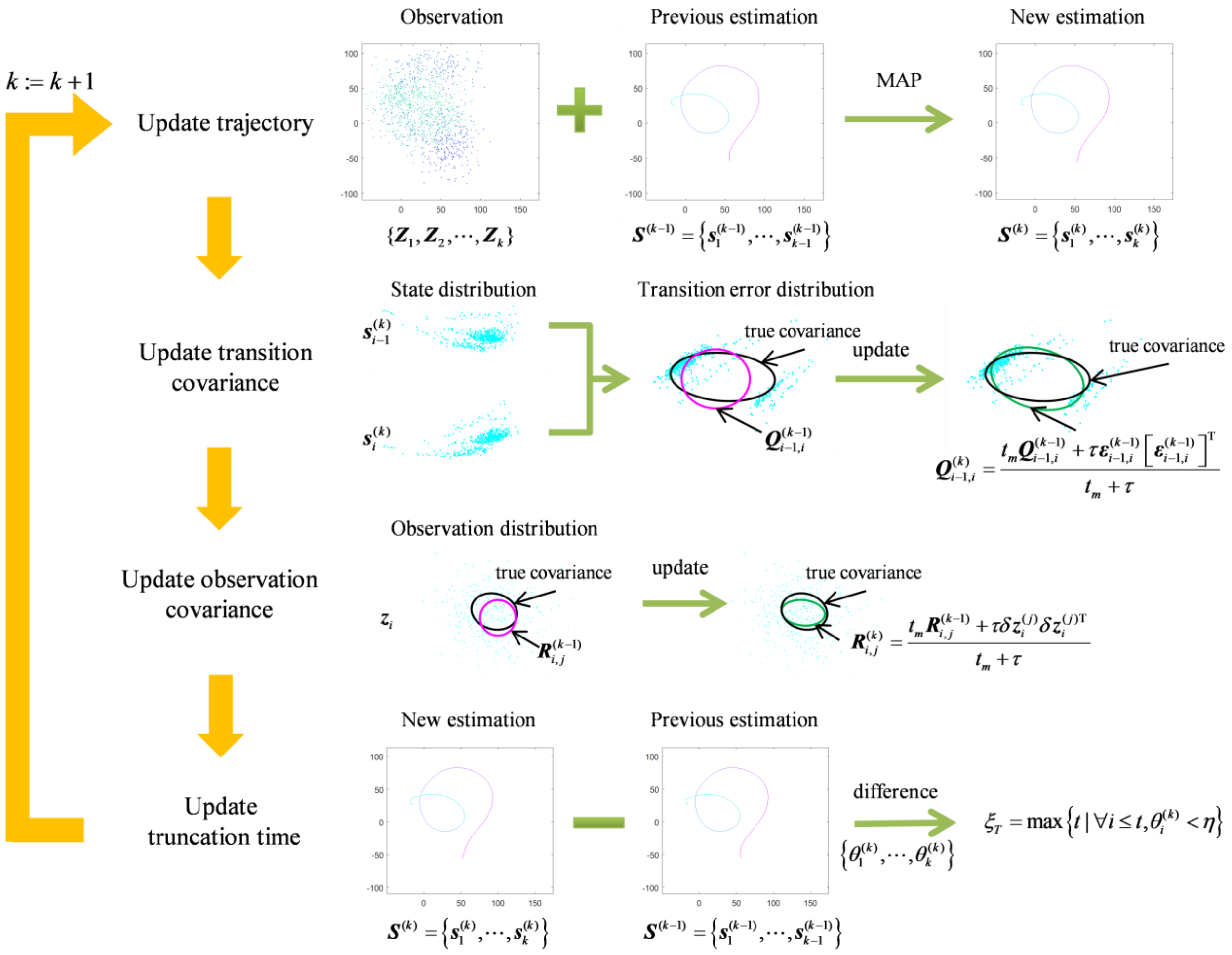

Fig. 1. The process of adaptive trajectory estimation. The first step is updating the trajectory based on the observations and previous estimation with sparse maximum a posteriori estimation (sparse MAP). The second step is updating the transition covariance of different times with a transition error. The third step is updating the observation covariance based on analysis of the observation error. Finally, the truncation time which confines the range to be updated will be worked out based on the difference between previous and updated trajectory.

Estimation algorithm is another emphasis. One famous approach is the Kalman filter (KF) [17] and Kalman smoother [7][10] for linear dynamic models. To expand the application to nonlinearity, the extended Kalman filter (EKF) and the Unscented Kalman Filter (UKF) [18][19] were developed. But above filters have common defects: a) They are sensitive to the model error and sensitive to the prior statistics; b) Fixed and zeros centered noise model is applied, which leads to the fragility to the observation drift. To overcome these problems, some specifically designed methods have been proposed, such as eXogenous Kalman Filter (XKF) [20], maximum correntropy Kalman filter (MCKF) [21], and the Robust Kalman Filtering (RKF) [22]. Each of these methods tries to solve certain problem, but none of them deals with the difficulties in a single light framework. Another famous method is the particle filter [23]-[26]. Through the resampling strategy [25], the particle distribution approximates the real state distribution. Compared with Kalman filters, it is much more robust to the biased prior statistics. However, because of the huge computation for maintaining a massive number of candidate trajectories, particle filter is difficult to be applied in real-time estimation.

Different from the filtering approaches, the optimization approaches can produce globally consistent result. A representative illustration is the graph optimization for SLAM [27][28]. However, the basic graph optimization is sensitive to dynamic model and prior statistics. If strongly nonlinear models are invoked with biased covariance matrices, wrong results and divergence may occur. Besides, sparse Cholesky factorization [27] and Krylov subspace methods [29] is usually applied to accelerate the calculation in graph optimization with sparse global correlations named loop closures (some states have direct relationship with the states long time ago). When being applied in a task with only regional correlations, they would bring a lot of unnecessary calculation. To taking advantage of the regional correlations, we have proposed a sparse maximum a posteriori estimation algorithm (sparse MAP) [8], in our previous work. It is efficient in short trajectory estimation. But the linear increasing calculation over time prevents the application to long term estimation.

Our motivation of designing estimation algorithm focus on improving the robustness to biased prior statistics while keeping efficiency by adaptively limits the updating range. Based on the previous work, sparse MAP [8], we propose a specifically designed optimization method called the adaptive trajectory estimation algorithm (AdaTE) with regional correlation hypothesis and PLS model. The process is plotted in Fig. 4. Firstly, the trajectory is updated based on the previous estimation and corresponding statistics. Then, the statistics (including transition covariance matrices and observation covariance matrices) is updated based on the newest trajectory and historical observations. Finally, the truncation time confining the range to be updated is calculated based on the difference between previous and updated trajectory. The motivations of these procedures are: a) The adaptation of transition covariance matrices improves the estimation robustness to the model error and prior statistics; b) The correction of observation covariance matrices improves the estimation robustness to the observation drift; c) The truncation time ensures the calculation is restricted to the necessary part, which avoids the linear improvement of calculation.

The contribution of this paper is threefold: a) A dynamic model based on Newton mechanics is proposed for non-cooperative maneuvering object, it sustains few parameters to describe the switch of typical states; b) The adaptive trajectory estimation algorithm is proposed to achieve robust and efficient trajectory estimation; c) Through the experiments, we show that the AdaTE can successfully estimate the trajectory in challenging situations and various backgrounds.

The remainder of this paper will be mentioned as follows. Section II familiarizes the reader with related researches in dynamic model and trajectory estimation algorithm. Section III presents the power limited steering model (PLS) and its convergence. Section IV presents the adaptive trajectory estimation algorithm. Section V introduces three experiments for AdaTE: trajectory estimation on 3D observations, Local navigation under random obstacle, trajectory optimization for visual tracking. The paper is summarized in Section VI.

## II. Related Works

Dynamic model and estimation algorithm are two key factors for trajectory estimation. In this section, we firstly introduce some subclass of dynamic model, their representative members, and their relationship and difference with PLS model. Then we talk about the estimation algorithm and make a contrast of filtering methods with optimization methods.

The assumptions in this paper are grouped as follow: 1) the dedicated dynamic model for particular object cannot be obtained; 2) all of the prior statistics, including transition covariance and observation covariance are inaccurate; 3) the correlation between states is regional, that means the state evolution is a Markov process; 4) part of the observations have drift.

### A. Dynamic Model

Some dedicated models [11] are designed for cooperative object and system. But they are restricted to dedicated machine under accessible control. For non-cooperative object whose control is unknown to the observer, it is hard to design a dedicated model. That means the trajectory of these objects cannot be accurately forecasted. A popular alternative is describing the movement as a stochastic process with random control. It is known as random model and can be classified into three subclasses: white noise models, Markov process models and Semi-Markov process models. One simplest and representative dynamic model is the Wiener-process acceleration model [2], where acceleration is assumed to obey the Wiener process. Many following constant acceleration (CA) models are inspired by this work.

Lets $\boldsymbol{x}$, $\boldsymbol{v}$ and $\boldsymbol{a}$ be the position, velocity and acceleration of the object, Setting the state vector as

$$\boldsymbol{s} = \begin{bmatrix} \boldsymbol{x} \\ \boldsymbol{v} \\ \boldsymbol{a} \end{bmatrix} \tag{1}$$

The Wiener-process acceleration model is presented as

$$\dot{\boldsymbol{s}} = \boldsymbol{A}\boldsymbol{s} + \boldsymbol{u} \tag{2}$$

where $\boldsymbol{u}$ is the control. It is an independent process (white noise) of acceleration with power spectral density $\boldsymbol{D}$. $\boldsymbol{A}$ is the state transition matrix,

$$\boldsymbol{A} = \begin{bmatrix} \mathbf{0} & \boldsymbol{I} & \mathbf{0} \\ \mathbf{0} & \mathbf{0} & \boldsymbol{I} \\ \mathbf{0} & \mathbf{0} & \mathbf{0} \end{bmatrix} \tag{3}$$

where $\boldsymbol{I}$ is the identity matrix.

Based on the antitype of Wiener-process, many following works have been proposed, such as the polynomial models, the Singer model [14], "current" model and semi-Markov jump process model [15]. Their improvement mainly focuses on the distribution of random vector, such as quantized levels of expectations,

$$\boldsymbol{u}(t) = \sum_{i=1}^{n} \bar{\boldsymbol{a}}_i P\{\boldsymbol{u}(t) = \bar{\boldsymbol{a}}_i \mid \boldsymbol{z}(s), s < t\} + \tilde{\boldsymbol{a}} \tag{4}$$

where $\bar{\boldsymbol{a}}_i$ is the quantized expectation of acceleration, $\tilde{\boldsymbol{a}}$ is the zero-mean random vector, and $\boldsymbol{z}(s), s < t$ is all of the observations before time $t$.

Another well-known approach is modifying the state transition matrix, such as adding the damping term

$$\boldsymbol{a} = -\beta \boldsymbol{v} + \boldsymbol{u} \tag{5}$$

where $\beta$ is the damping factor.

Whatever they are modified, the dynamic models are integration of many linear models for different status. If they want to distinguish different typical movements in one model, they need a lot of quantized levels and corresponding conditional probability based on reliable prior information. A typical illustration is the quantized expectations of acceleration $\bar{\boldsymbol{a}}_i, i = 1, 2, \cdots, n$ defined in (4). Thus unreliable prior information is a challenge for them.

Another impressive work is the generative model for maneuvering target tracking (GMTM) [30], where the motive force is divided into the axial force and centripetal force to describe axial acceleration and angular velocity based on Newton mechanics. But the problem is that, to limit the output power given by the product of force with speed, GMTM models the speed-force constraints using conditional Rayleigh distributions. It increases a lot of calculation.

To inherit the advantage and remedy deficiency of GMTM [30], we propose the power-limited steering model (PLS). Its main innovation is the combination of instantaneous power theory and instantaneous angular velocity theory from Newton mechanics. Compared with GMTM, PLS is a lot of cheaper in limiting the output power through the power damping interaction. Moreover, PLS is a nonlinear model, thus it needs fewer prior parameters to describe the typical states compared with Markov process models. However, the nonlinearity of PLS leads to the extra difficulty in deriving corresponding discrete time model. It will be specifically discussed in next section of this paper.

### *B. Estimation Algorithm*

The estimation algorithm can be conventionally separated into two categories: filtering approaches and optimization approaches. Normally, filtering approaches result in the expectation of the states at each time under certain distribution, while optimization approaches estimate a trajectory with maximum global probability. The filtering approaches were widely used at early time because they were believed to require fewer computations compared with optimization approaches. Until the optimization framework was found to have sparse associations, it starts to be popular.

A well-known filtering category is the Kalman Filter family. The member includes Kalman Filter (KF), extended Kalman filter (EKF), Unscented Kalman Filter (UKF) [18][19], Cubature Kalman Filters (CKF) [31][32], and some recent developed methods such as eXogenous Kalman Filter (XKF) [20] and the Maximum Correntropy Kalman filter (MCKF) [21]. For the real-time state estimation, they execute a forward propagation:

$$\begin{aligned} &\mathrm{E}[\boldsymbol{s}_k \mid \mathcal{Z}_k] = \\ &\int \left[ \int \frac{P(\boldsymbol{Z}_k \mid \boldsymbol{s}_k) P(\boldsymbol{s}_k \mid \boldsymbol{s}_{k-1}, \mathcal{Z}_{k-1})}{P(\boldsymbol{Z}_k)} d\boldsymbol{s}_{k-1} \right] \boldsymbol{s}_k d\boldsymbol{s}_k \end{aligned} \tag{6}$$

where $\boldsymbol{s}_i, i \in \{1, \cdots, k\}$ is the state at time $i$, $\mathcal{Z}_k$ is the observations at time $k$ which might contain several independent observations, and $\mathcal{Z}_k = \{\boldsymbol{Z}_1, \cdots, \boldsymbol{Z}_k\}$ is the set of observations from time 1 to time $k$.

In trajectory estimation, they are known as smoothers to correct the previous states in an inverse time order:

$$\begin{aligned} &\mathrm{E}[\boldsymbol{s}_i \mid \mathcal{Z}_k] = \int \left[ \int P(\boldsymbol{s}_{i+1} \mid \boldsymbol{s}_i, \mathcal{Z}_k) P(\boldsymbol{s}_i \mid \mathcal{Z}_{k-1}) d\boldsymbol{s}_{i+1} \right] \boldsymbol{s}_i d\boldsymbol{s}_i, \\ &i = k-1, \cdots, 1 \end{aligned} \tag{7}$$

In (7), trajectory is estimated as a set of expectations at each moment. The result is not the trajectory with the highest probability but rather a compromise of possibility. If there is a dedicated dynamic model and accurate prior statistics (especially the conditional distribution), it would be easy to obtain an accurate trajectory; otherwise if parts of the priori statistics are inaccurate or even biased, there will be systematic deviation of result. For example, the prior information indicates that the noise of observation is zero-mean while it has drift in fact, the estimated trajectory will shift to one side. Besides, how to set the truncation time to limit the updating range of trajectory is another problem for the filtering approaches. Although some researches worked on this issue, like restricted memory filtering, they still left some problem such as how to adjust the length of memory online.

Different from the filtering approaches, the optimization approaches, such as graph optimization [33]-[35] for SLAM [4]-[6], can be can be mentioned as finding the trajectory with the highest probability from observations:

$$\mathcal{S}_k = \arg\max P(\mathcal{S}_k \mid \mathcal{Z}_k) \tag{8}$$

where $\mathcal{S}_k = \{\boldsymbol{s}_1, \cdots, \boldsymbol{s}_k\}$ is a set of states from time 1 to time *k*.

For a linear model with a simple distribution, (8) has an analytic solution. For nonlinear model or complex distribution that do not have analytic solutions, numerical solutions can be worked out through the Newton-Raphson, BFGS [36] or the gradient descent algorithm [37].

In general, the optimization approaches should have sequential iterative algorithm for online estimation. Because the last estimation is optimized based on the best estimation in the previous time, and the initial estimation has a smooth and small solution space, the result of sequential iteration will be always close to the global optimum. It should be noticed that, if a new observation is inconsistent with the previous estimation, the optimization algorithm often needs extra time to correct the trajectory. Otherwise if we choose to believe the previous estimation, to what extent and on what time we should correct the trajectory? Moreover, sparse Cholesky factorization and Krylov subspace methods [29] are usually applied to accelerate the optimization process [27]. They get excellent performance in optimization with global correlations. Although some outstanding works [29] have reduced their complexity, the direct application in a task only with regional correlations (such as visual tracking) still brings a lot of unnecessary computation.

Both of the listed filtering approaches and optimization approaches have a common problem: they urgently depend on prior statistics. If the prior statistics are incorrect, the result of these algorithms may have extra error. Moreover, if the dynamic model is nonlinear, such as the PLS in this paper, this error may be magnified.

To overcome the problems above, the newly designed AdaTE should have following abilities: (a) be insensitive to prior statistics, (b) can find out the observation outliners and correct their statistics, (c) adaptively limits the updating length of trajectory based on the estimation fluctuation.

Dynamic model is the foundation of trajectory estimation. Before the introduction of AdaTE, we should describe the motion of object with a dynamic model at first.

## III. Power Limited Steering Model

In some dynamics models, the movements in orthogonal directions are assumed to be uncoupled with each other [38][39]. But in fact the opposite is true, and the assumption will weaken the ability of trajectory prediction in many cases like making a turn. We learn from the previous works and propose the power-limited steering model (PLS). It is a nonlinear model based on a natural combination of instantaneous power theory and instantaneous angular velocity theory from Newtonian mechanics. In general, PLS obeys two design philosophies: a) it should overcome the problem of infinite speed in prediction in simplest way; b) it should have fewer prior parameters to describe the typical states compared with Markov process models. Firstly, we will introduce the PLS in continuous time. Then we will derive the corresponding form in discrete time.

### A. Power-Limited Steering Model in Continuous Time

Let $\boldsymbol{x}$, $\boldsymbol{v}$ and $\boldsymbol{a}$ be the position, velocity and acceleration of the object The movement of an object can be decomposed into the axial and transverse direction. According to Newtonian mechanics, if the object is driven by power $P$ and receives resistance $\beta$ and damping $\alpha$ , the axial acceleration is:

$$\boldsymbol{a}_A = -\alpha\boldsymbol{v} - \beta\frac{\boldsymbol{v}}{\|\boldsymbol{v}\|} + \frac{P\boldsymbol{v}}{m\|\boldsymbol{v}\|^2}. \tag{9}$$

where $\boldsymbol{a}_A$ is the vector of axial acceleration, $\boldsymbol{v}$ is the vector of velocity, $\|\boldsymbol{v}\|$ is the norm of $\boldsymbol{v}$ , $m$ is the mass of object.

Setting $p = \frac{P}{m}$ be the power-mass ratio (specific power), which represents the power assigned to a unit mass in physics, (9) can be rewritten as

$$\boldsymbol{a}_A = -\alpha\boldsymbol{v} - \beta\frac{\boldsymbol{v}}{\|\boldsymbol{v}\|} + p\frac{\boldsymbol{v}}{\|\boldsymbol{v}\|^2} \tag{10}$$

It is easy to find that when the power is a constant positive number, the velocity will be asymptotically stabilized to $\frac{-\beta + \sqrt{\beta^2 + 4\alpha p}}{2\alpha}$.

Similar to (10), the acceleration in the transverse direction is given by

$$\boldsymbol{a}_T = \boldsymbol{c}_T \times \boldsymbol{v} = [\boldsymbol{c}_T]_\times \boldsymbol{v} \tag{11}$$

which is orthogonal to $\boldsymbol{v}$ , where vector $\boldsymbol{c}_T$ is the instantaneous angular velocity.

The variation of $p$ and $\boldsymbol{c}_T$ should be smooth, so we set $\dot{p} = u_A$ and $\dot{\boldsymbol{c}}_T = \boldsymbol{u}_T$ as the axial and transverse controls. As a maneuvering object, the controls have high randomness and low temporal correlation. So $u_A$ and $\boldsymbol{u}_T$ are modeled as time-independent and zero-mean random pulses obeying the normal distribution,

$$u_A(t) \sim \mathcal{N}\{0, \delta_t(s)D_A\} \tag{12}$$

$$\boldsymbol{u}_T(t) \sim \mathcal{N}\{\boldsymbol{0}, \delta_t(s)\boldsymbol{D}_T\} \tag{13}$$

Where $\delta_t(s)$ is the delta function, $D_A$ and $\boldsymbol{D}_T$ are the power spectral density of $u_A$ and $\boldsymbol{u}_T$ ..

Combining the axial and the transverse movement, the power-limited steering model in continuous time satisfies

$$\begin{cases} \dot{\boldsymbol{x}} = \boldsymbol{v} \\ \dot{\boldsymbol{v}} = -\alpha \boldsymbol{v} - \beta \dfrac{\boldsymbol{v}}{\|\boldsymbol{v}\|} + p \dfrac{\boldsymbol{v}}{\|\boldsymbol{v}\|^2} - [\boldsymbol{v}]_\times \boldsymbol{c}_T \\ \dot{p} = u_A \\ \dot{\boldsymbol{c}}_T = \boldsymbol{u}_T \end{cases} \tag{14}$$

Next we will discuss the solution of (14).

### B. The Short-Term Evolution Of Velocity

Noticing (14) is a nonlinear stochastic function, where $u_A$, $\boldsymbol{u}_T$ will cause the perturbation of $p$ and $\boldsymbol{c}_T$, which is hard to be solved directly. To simplify the problem, we will solve the velocity ignoring the perturbation of $p$ and $\boldsymbol{c}_T$ at first, and then compensate the effect of perturbation.

In differential equation (14), the axial and transverse velocity can be orthogonally separated into two groups:

$$\begin{cases} \|\boldsymbol{v}\| \dfrac{d\|\boldsymbol{v}\|}{dt} = p - \alpha \|\boldsymbol{v}\|^2 - \beta \|\boldsymbol{v}\| \\ \dot{p} = u_A \end{cases} \tag{15}$$

and

$$\begin{cases} \dfrac{d}{dt} \dfrac{\boldsymbol{v}}{\|\boldsymbol{v}\|} = [\boldsymbol{c}_T]_\times \dfrac{\boldsymbol{v}}{\|\boldsymbol{v}\|} \\ \dot{\boldsymbol{c}}_T = \boldsymbol{u}_T \end{cases} \tag{16}$$

where $u_A$ and $\boldsymbol{u}_T$ obeys (12) and (13).

Firstly, ignoring the perturbations of $p$ and $\boldsymbol{c}_T$ during a short time $\tau$, we can solve the biased prediction of velocity $\hat{\boldsymbol{v}}(t+\tau)$ from appendix A:

$$\begin{cases} \left( \dfrac{\|\hat{\boldsymbol{v}}(t+\tau)\| + \varsigma_1}{\|\boldsymbol{v}(t)\| + \varsigma_1} \right)^{\varsigma_1} \left| \dfrac{\|\hat{\boldsymbol{v}}(t+\tau)\| + \varsigma_2}{\|\boldsymbol{v}(t)\| + \varsigma_2} \right|^{-\varsigma_2} = e^{-\gamma\tau} \\ \dfrac{\hat{\boldsymbol{v}}(t+\tau)}{\|\hat{\boldsymbol{v}}(t+\tau)\|} = \exp\left(\tau [\boldsymbol{c}_{\mathrm{T}}(t)]_\times\right) \dfrac{\boldsymbol{v}(t)}{\|\boldsymbol{v}(t)\|} \end{cases} \tag{17}$$

where $\varsigma_1$ and $\varsigma_2$ are defined in (48), and $\gamma$ is defined in (49).

It is hard to solve the analytical solution of $\hat{\boldsymbol{v}}(t+\tau)$ from (17) directly, but we can quickly work out the arithmetic solution from there with Newton-Raphson method.

After the estimation of $\hat{\boldsymbol{v}}(t+\tau)$ with perturbation ignorance, the compensation of perturbation should be added to the estimated axial velocity. Based on the renormalization group, the compensation of axial velocity during time $(t, t+\tau)$ can be derived from appendix B,

$$\Delta\upsilon(t,t+\tau) = -\frac{\upsilon_0 D_A}{2\mu(t)} \left[ \tau - \frac{2\upsilon_0^2}{\mu(t)} \log \frac{\upsilon_0^2 + \tau\mu(t)}{\upsilon_0^2} + \frac{\upsilon_0^2 \tau}{\upsilon_0^2 + \tau\mu(t)} \right] \tag{18}$$

where $\Delta\upsilon(t,t+\tau)$ is the compensation of axial velocity, $\mu(t) = p(t) - \upsilon_0(\alpha\upsilon_0 + \beta)$, $\upsilon_0 = \|\boldsymbol{v}(t)\|$.

Combining (17) and (18), we get the corrected prediction of axial velocity,

$$\|\bar{\boldsymbol{v}}(t+\tau)\| = \|\hat{\boldsymbol{v}}(t+\tau)\| + \Delta\upsilon(t,t+\tau) \tag{19}$$

and the direction of velocity,

$$\frac{\bar{\boldsymbol{v}}(t+\tau)}{\|\bar{\boldsymbol{v}}(t+\tau)\|} = \exp\left(\tau [\boldsymbol{c}_{\mathrm{T}}(t)]_\times\right) \frac{\boldsymbol{v}(t)}{\|\boldsymbol{v}(t)\|} \tag{20}$$

### C. PLS Model in Discrete Time

To match the requirement of the optimization algorithm in next section, the state transition should be written in a linear function in discrete time

$$\boldsymbol{s}_{k+1} = \boldsymbol{\Phi}_{k,k+1} \boldsymbol{s}_k + \boldsymbol{\varepsilon}_{k,k+1} \tag{21}$$

where $\boldsymbol{s}=\left[\boldsymbol{x}^{\mathrm{T}}\ \boldsymbol{v}^{\mathrm{T}}\ \ p\ \ \boldsymbol{c}_T{}^{\mathrm{T}}\right]^{\mathrm{T}}$, $\boldsymbol{s}_k$ and $\boldsymbol{s}_{k+1}$ is the state vector at time $k$ and $k+1$.

$\boldsymbol{\varepsilon}_{k,k+1}$ is the transition error, its integral form in continuous time is,

$$\boldsymbol{\varepsilon}(t,t+\tau)=\int_0^{\tau}\boldsymbol{\Phi}(t+s,t+\tau)\boldsymbol{u}(t+s)ds \tag{22}$$

where $\boldsymbol{u}=[\mathbf{0}\quad \mathbf{0}\quad u_A^{\mathrm{T}}\quad \boldsymbol{u}_T^{\mathrm{T}}]^{\mathrm{T}}$ is the control vector, $u_A$ and $\boldsymbol{u}_T$ is defined in (12) and (13).

To work out the $\boldsymbol{\Phi}_{k,k+1}$ and $\boldsymbol{\varepsilon}_{k,k+1}$ from velocity evolution (19)-(20), the state transition in continuous time $\boldsymbol{\Phi}(t,t+\tau)$ should be solved at first.

Form previous section, the evolution of velocity satisfies:

$$\boldsymbol{v}(t+\tau)=\bar{\boldsymbol{v}}(t+\tau)+\delta\boldsymbol{v}(t+\tau) \tag{23}$$

where $\bar{\boldsymbol{v}}(t+\tau)$ is the predicted velocity from (19)-(20).

From appendix C, after expanding the $\boldsymbol{v}(t+\tau)$ into a linear function of $\boldsymbol{v}(t)$, $p(t)$ and $\boldsymbol{c}_T(t)$, each submatrix of transition matrix

$$\boldsymbol{\Phi}(t,t+\tau)=\begin{bmatrix}\boldsymbol{I} & \boldsymbol{\Phi}_{12}(t,t+\tau) & \boldsymbol{\Phi}_{13}(t,t+\tau) & \boldsymbol{\Phi}_{14}(t,t+\tau)\\ \mathbf{0} & \boldsymbol{\Phi}_{22}(t,t+\tau) & \boldsymbol{\Phi}_{23}(t,t+\tau) & \boldsymbol{\Phi}_{24}(t,t+\tau)\\ \mathbf{0} & \mathbf{0} & 1 & \mathbf{0}\\ \mathbf{0} & \mathbf{0} & \mathbf{0} & \boldsymbol{I}\end{bmatrix} \tag{24}$$

can be obtained from (70)-(71), where $\boldsymbol{I}$ is the identity matrix

With the transition matrix, we can analyze the transition error defined in (22).

Because the expectation of velocity perturbation has been compensated in (19), $\boldsymbol{\varepsilon}(t,t+\tau)$ will be a zero-mean random vector which satisfies $\boldsymbol{\varepsilon}(t,t+\tau)\sim\mathcal{N}\left\{\mathbf{0},\boldsymbol{Q}(t,t+\tau)\right\}$. So we only need to work out the transition covariance of $\boldsymbol{\varepsilon}(t,t+\tau)$ from appendix D.

Let $\boldsymbol{\Phi}_{k,k+1}=\boldsymbol{\Phi}(t_k,t_{k+1})$, and $\boldsymbol{Q}_{k,k+1}=\boldsymbol{Q}(t_k,t_{k+1})$, the state transition in discrete time described in (21) can be obtained.

It should be noted that although $\boldsymbol{\Phi}_{k,k+1}$ implicitly includes the state vector, and the calculation is approximate, the estimation algorithm based on PLS still converges. This is because the transition matrix $\boldsymbol{\Phi}_{k,k+1}$ is adaptively corrected by estimation algorithm in section IV. It can be verified by the experiments in section V.

### *D. Convergence of PLS*

The axial velocity defined in (15) can be rewritten as:

$$\frac{d\|\boldsymbol{v}\|}{dt}=-\frac{(\|\boldsymbol{v}\|+\varsigma_1)(\|\boldsymbol{v}\|+\varsigma_2)}{\|\boldsymbol{v}\|} \tag{25}$$

where $\varsigma_1$ and $\varsigma_2$ are defined in (47).

The convergence is discussed in cases.

(a) $p\geq 0$

Due to $\|\boldsymbol{v}\|+\varsigma_1>\|\boldsymbol{v}\|\geq 0$, in this situation, the axial acceleration satisfies,

$$\begin{cases}\dfrac{d\|\boldsymbol{v}\|}{dt}<-(\|\boldsymbol{v}\|+\varsigma_2)<0,\|\boldsymbol{v}\|>-\varsigma_2\\ \dfrac{d\|\boldsymbol{v}\|}{dt}>-(\|\boldsymbol{v}\|+\varsigma_2)>0,\|\boldsymbol{v}\|<-\varsigma_2\end{cases} \tag{26}$$

It means when $\|\boldsymbol{v}\|>-\varsigma_2$, $\|\boldsymbol{v}\|$ will be decrease; when $\|\boldsymbol{v}\|<-\varsigma_2$, $\|\boldsymbol{v}\|$ will be increase. Thus $\|\boldsymbol{v}\|$ global convergence to $\varsigma_2$.

(b) $p<0$

In this situation, because $\|\boldsymbol{v}\|\geq 0$ always stand up, there is $p-\alpha\|\boldsymbol{v}\|^2-\beta\|\boldsymbol{v}\|<0$, so $\dfrac{d\|\boldsymbol{v}\|}{dt}<0$, and the object decelerate. When $\|\boldsymbol{v}\|=0$, the deceleration process stop.

It is important to note that when $\|\boldsymbol{v}(t)\|=0$, although the item $p/\|\boldsymbol{v}(t)\|$ in axial acceleration is infinity in an instant, but the integral of acceleration during a period of time is finite. The reason is that instant infinity acceleration will give the speed a non-zero value, then the $p/\|\boldsymbol{v}(t)\|$ will restricted to finite value as a feedback. It can be justified by taking $\|\boldsymbol{v}(t)\|=0$ into (17), $\|\hat{\boldsymbol{v}}(t+\tau)\|$ is not infinity.

Compared with that the GMTM applying conditional Rayleigh distributions for the convergence of speed, PLS is obviously simpler in realizing the convergence.

To obtain a trajectory from observations, we not only need the dynamic model, but also a precise and robust estimation algorithm. Next section will introduce how to design this algorithm.

## IV. Adaptive Trajectory Estimation

As a maneuvering object, the prior statistics are often inaccurate. This will lead to the extra error of the algorithms which urgently relying on the prior statistics. While the dynamic model is nonlinear, the error may be magnified. Comprehensively considering other problems, such as the observations drift, the adaptive trajectory estimation algorithm should have following abilities: (a) be insensitive to prior statistics, (b) can find out the outstanding observations and correct their statistics, (c) adjust the truncation time based on the estimation fluctuation. According to these principles, we design the AdaTE for online trajectory estimation. In general, it contains 4 steps: (a) updating the trajectory with sparse maximum a posterior estimation (sparse MAP); (b) updating the transition covariance of different times with a transition error; (c) updating the observation covariance based on analysis of the observation error; (d) update the truncation time based on the difference between previous and latest trajectory.

### *A. Updating the Trajectory with Sparse Map*

Setting $\xi_T$ be the truncation time, $\mathcal{S}_k=\{\boldsymbol{s}_1,\cdots,\boldsymbol{s}_k\}$ be the trajectory of states at time $k$, $\mathcal{Z}_k=\{\boldsymbol{Z}_1,\cdots,\boldsymbol{Z}_k\}$ be the set of observations from time 1 to time $k$, the updated trajectory under optimization framework satisfies

$$\mathcal{S}_k^{(\xi_T:k)}=\arg\max P(\mathcal{S}_k^{(\xi_T:k)}\mid\mathcal{S}_{k-1},\mathcal{Z}_k) \tag{27}$$

where $\mathcal{S}_k^{(\xi_T:k)}\subseteq\mathcal{S}_k$ is the part of trajectory during time $[\xi_T,k]$.

Assuming the observers are independent of each other, the probability of trajectory is

$$P(\mathcal{S}_k\mid\mathcal{Z}_k)=P(\boldsymbol{s}_1)\prod_{j=1}^{n}\frac{P(\boldsymbol{z}_1^{(j)}\mid\boldsymbol{s}_1)}{P(\boldsymbol{z}_1^{(j)})}\left\{\prod_{i=2}^{k}P(\boldsymbol{s}_i\mid\boldsymbol{s}_{i-1})\prod_{j=1}^{n}\frac{P(\boldsymbol{z}_i^{(j)}\mid\boldsymbol{s}_i)}{P(\boldsymbol{z}_i^{(j)})}\right\} \tag{28}$$

Typically, the observable is the position of object. It is linear with the state:

$$\boldsymbol{z}=\boldsymbol{H}\boldsymbol{s}+\delta\boldsymbol{z} \tag{29}$$

where $\boldsymbol{H}$ is the observation matrix, $\delta\boldsymbol{z}$ is the observation error temporary assumed to obey normal distribution $\mathcal{N}\{\boldsymbol{0},\boldsymbol{R}\}$, where $\boldsymbol{R}$ has been updated in previous estimation.

Assuming the state transition satisfies the PLS model (21), where transition matrix $\boldsymbol{\Phi}_{i,i+1}, i=1,2,\cdots,k-1$ is in the form of (24) and calculated on the previous trajectory $\mathcal{S}_{k-1}$. And each conditional probability in (28) satisfies:

$$\begin{cases}\log P(\boldsymbol{s}_i\mid\boldsymbol{s}_{i-1})=-\dfrac{1}{2}\left\|\boldsymbol{\Phi}_{i-1,i}\boldsymbol{s}_{i-1}-\boldsymbol{s}_i\right\|^2_{\boldsymbol{Q}^{-1}_{i-1,i}}\\ \log P(\boldsymbol{z}_i^{(j)}\mid\boldsymbol{s}_i)=-\dfrac{1}{2}\left\|\boldsymbol{H}\boldsymbol{s}_i-\boldsymbol{z}_i^{(j)}\right\|^2_{\boldsymbol{R}^{-1}_{i,j}}\end{cases} \tag{30}$$

Maximizing probability (28), the optimal trajectory satisfies the sparse linear equations

$$\mathcal{M}_k^{(\xi_T:k)}\mathcal{S}_k^{(\xi_T:k)}=\mathcal{B}_k^{(\xi_T:k)} \tag{31}$$

where $\mathcal{M}_k^{(\xi_T:k)}=\{\boldsymbol{M}_{i,j}, i,j=\xi_T,\cdots,k\}$ is a sparse matrix, whose nonzero elements satisfies:

$$\boldsymbol{M}_{i,i}=\begin{cases}\boldsymbol{H}^{\mathrm{T}}\sum_{j=1}^{n}\boldsymbol{R}^{-1}_{i,j}\boldsymbol{H}+\boldsymbol{\Phi}^{\mathrm{T}}_{i,i+1}\boldsymbol{Q}^{-1}_{i,i+1}\boldsymbol{\Phi}_{i,i+1}+\boldsymbol{Q}^{-1}_{i-1,i}, & i=\xi_T,\cdots,k-1\\ \boldsymbol{H}^{\mathrm{T}}\sum_{j=1}^{n}\boldsymbol{R}^{-1}_{k,j}\boldsymbol{H}+\boldsymbol{Q}^{-1}_{k-1,k}, & i=k\end{cases} \tag{32}$$

$$\boldsymbol{M}_{i,i-1}=\boldsymbol{M}^{\mathrm{T}}_{i-1,i}=\boldsymbol{Q}^{-1}_{i-1,i}\boldsymbol{\Phi}_{i-1,i},\quad i=\xi_T+1,\cdots,k \tag{33}$$

$\mathcal{B}_k^{(\xi_T:k)}=\{\boldsymbol{b}_i, i=\xi_T,\cdots,k\}$ is the set of vectors, the elements are:

$$\boldsymbol{b}_i = \begin{cases} \boldsymbol{H}^{\mathrm{T}} \sum_{j=1}^{n} \boldsymbol{R}_{1,j}^{-1} \boldsymbol{z}_1^{(j)} + \boldsymbol{Q}_{i-1,i}^{-1} \boldsymbol{\Phi}_{i-1,i} \boldsymbol{x}_{i-1}, i = \xi_T \\ \boldsymbol{H}^{\mathrm{T}} \sum_{j=1}^{m} \boldsymbol{R}_{i,j}^{-1} \boldsymbol{z}_i^{(j)}, i = \xi_T + 1, \cdots, k \end{cases} \tag{34}$$

Noticing that the set $\mathcal{M}_k^{(\xi_T:k)}$ is a band-diagonal symmetric matrix, equation (31) can be rapidly solved.

### B. Adaptation of Transition Covariance

As a maneuvering object, it is hard to get an accurate prior statistics. This will lead to the extra error of the algorithms which urgently relying on the prior statistics. While the PLS is a nonlinear model, the error may be magnified. To be insensitive to the prior statistics, the transition covariance can be updated by a fading memory strategy,

$$\boldsymbol{Q}_{i,i+1} := \frac{t_m \boldsymbol{Q}_{i,i+1} + \tau \boldsymbol{\varepsilon}_{i,i+1}^{(k+1)} \left[ \boldsymbol{\varepsilon}_{i,i+1}^{(k+1)} \right]^{\mathrm{T}}}{t_m + \tau}, i = 1, 2, \cdots, k-1 \tag{35}$$

where $\boldsymbol{Q}_{i,i+1}$ is the transition covariance, respectively, and $t_m$ is an empirical hyper-parameter of the memory time. $\boldsymbol{\varepsilon}_{i,i+1}^{(k+1)}$ is the transition error defined as:

$$\boldsymbol{\varepsilon}_{i,i+1}^{(k+1)} = \boldsymbol{s}_{i+1}^{(k+1)} - \boldsymbol{\Phi}_{k,k+1} \boldsymbol{s}_i^{(k+1)},, i = 1, 2, \cdots, k-1 \tag{36}$$

where $\boldsymbol{s}_i^{(k+1)}$ and $\boldsymbol{s}_{i+1}^{(k+1)}$ are the MAP of trajectory at time $k+1$.

Through updating the transition covariance, the overestimated part of the trajectory in maneuvering can be revised. It improves the robustness to the biased prior statistics.

### C. Adaptation of Observation Covariance

In most of the observation models, the prior statistics indicate that the noise of observation is zero-mean while it has drift in fact. This will lead to the systematic deviation of estimation. To solve the problem, we use an adaptation strategy to correct the covariance of abnormal observations based on the difference between the real observation and the estimated one from trajectory.

First, we define the observation bias:

$$\delta \boldsymbol{z}_i^{(j)} = \boldsymbol{z}_i^{(j)} - \boldsymbol{H}\boldsymbol{s}_i \tag{37}$$

And the corresponding deviation:

$$e_z(\boldsymbol{z}_i^{(j)}) = \left\| \delta \boldsymbol{z}_i^{(j)} \right\|_{\boldsymbol{R}_{i,j}^{-1}} \tag{38}$$

where $\boldsymbol{z}_i^{(j)}$ is the real observation, $\boldsymbol{H}\boldsymbol{s}_i$ is the estimated observation from trajectory.

Then, the observation whose deviation is greater than 3 times of their RMSE will be considered as abnormal observation, and its covariance will be corrected with

$$\boldsymbol{R}_{i,j} = \frac{t_m \boldsymbol{R}_{i,j} + \tau \delta \boldsymbol{z}_i^{(j)} \delta \boldsymbol{z}_i^{(j)\mathrm{T}}}{t_m + \tau} \tag{39}$$

Note that under the optimization framework with multiple sensors, not only the covariance of significantly deviated observations but also the covariance of drift observations are corrected in (39). Compared with classical filtering methods, it effectively overcomes the observation drift.

### D. Truncation Time

While the trajectory only has regional correlations, the correlation between the observation and the state is inversed to the time interval. This means the latest observations do not relate to the early states. In another word, some earlier states do not need to be updated. Based on this phenomenon, we defined a truncation time to limit the range of trajectory updating. It avoids the linear improvement of calculation in trajectory estimation.

The truncation index is defined as a weighted norm of difference vector between the same states in neighboring time:

$$\theta_i^{(k)} = \left\| \boldsymbol{s}_i^{(k)} - \boldsymbol{s}_i^{(k-1)} \right\|_{\psi}, i = 2, 3, \cdots, k \tag{40}$$

where $\boldsymbol{s}_i^{(k)}$ and $\boldsymbol{s}_i^{(k-1)}$ are the *i*-th state of the trajectory estimated at time $k$ and time $k-1$, and $\psi$ is a diagonal matrix quantize the weight of each dimension of $\boldsymbol{s}$.

The truncation time is defined as the last moment when all of previous truncation index is smaller than the threshold.

$$\xi_T = \max \left\{ t \mid \forall i \le t, \theta_i^{(k)} < \eta \right\} \tag{41}$$

where $\eta$ is the threshold of truncation index.

If $\xi_T > 0$, the part of trajectory before $\xi_T$ will be frozen, and only the part after that will be updated. To reserve sufficient observations for the trajectory estimation, the truncation time is bounded by $\xi_T \le k-10$.

### E. Adaptive Trajectory Estimation Algorithm

Summarizing the above steps, the procedures of AdaTE is achieved in Algorithm.

To analyze the real-time performance of AdaTE, we compare the time consumptions of AdaTE algorithm with a naive sparse MAP algorithm [8] in a dataset containing 600 observations. Fig. 2 shows the time consumption of two algorithms in trajectory estimation. The result is obvious: compared with the sparse MAP, the AdaTE successfully limits the time consumption.

Then we will verify the convergence of AdaTE through the experiments in section V.

---

**Algorithm:** Adaptive trajectory estimation

---

**Input**: Estimation at previous moment $\mathcal{S}_{k-1}$ ; parameters $\xi_T$ , $D_A$ , $\boldsymbol{D}_T$ , $\boldsymbol{Q}_{i-1,i}$, $\boldsymbol{R}_{i,j}$, $i \in \{1,\cdots,k-1\}$ ; $\boldsymbol{Z}_i, i \in \{1,\cdots,k\}$ .

**Output**: Updated trajectory $\mathcal{S}_k$ ; parameters and statistics $\xi_T$ , $\boldsymbol{Q}_{i-1,i}, \boldsymbol{R}_{i,j}; i \in \{1,\cdots,k\}$

1: **FOR** $i = \xi_T,\cdots,k-1$

2: Update transition matrices $\boldsymbol{\Phi}_{i-1,i}$ with (63)-(71).

3: **END FOR**

4: Calculate the last transition covariance $\boldsymbol{Q}_{k-1,k}$ with (74)-(79).

5: Calculate the coefficients $\mathcal{M}_k^{(\xi_T:k)}$ and $\mathcal{B}_k^{(\xi_T:k)}$ with (32)-(34).

6: Update the part of trajectory $\mathcal{S}_k^{(\xi_T:k)}$ through solving equation (31); then combining updated part with previous part $\mathcal{S}_k = \{\mathcal{S}_{k-1}^{(1:\xi_T-1)}, \mathcal{S}_k^{(\xi_T:k)}\}$ .

7: **FOR** $i = \xi_T,\cdots,k-1$

8: Update the transition covariance $\boldsymbol{Q}_{i-1,i}$ with (35)-(36).

9: Update the observation covariance $\boldsymbol{R}_{i,j}$ with (37)-(39).

10: **END FOR**

11: Update the truncation time $\xi_T$ through (40)-(41)

---

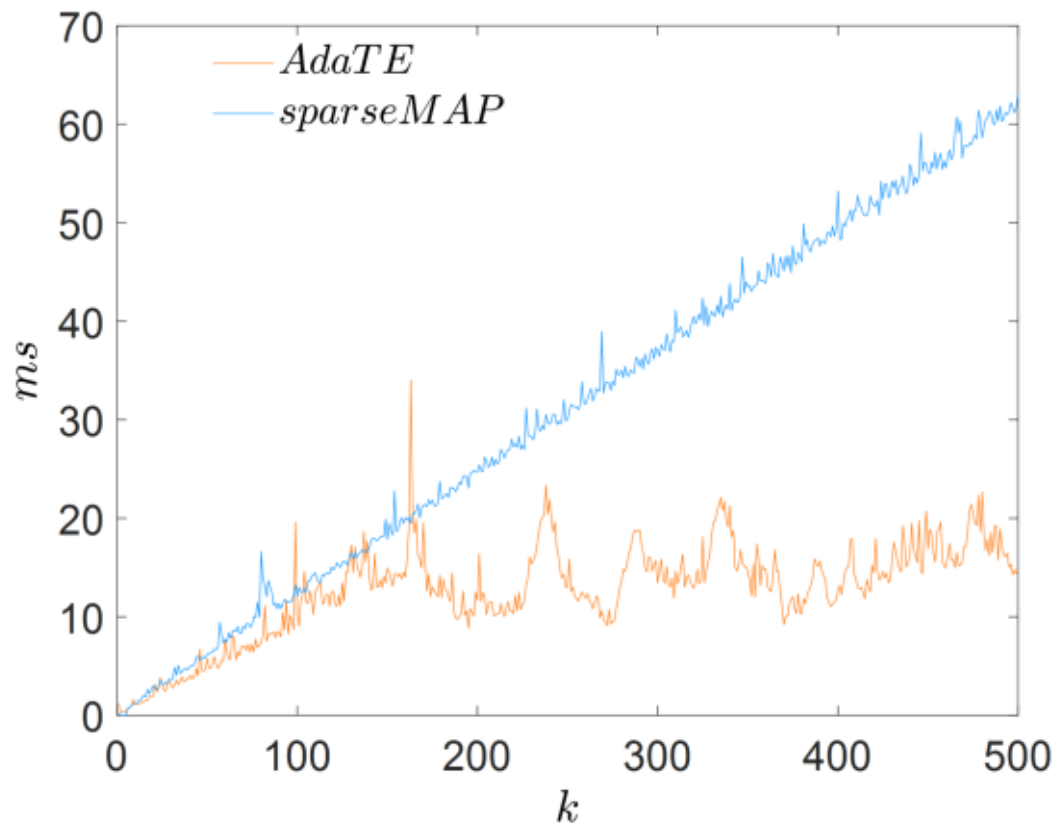

Fig. 2. The efficiency of AdaTE and naive sparse MAP. The time consumption of sparse MAP linearly increases with the number of observations, while the time consumption of AdaTE fluctuates around 17 ms ($\eta$ is set to 0.00001, number of observer is set to one).

## V. Experiments

To evaluation the convergence and performance of AdaTE, we conducted three experiments: (a) typical trajectory estimation on 3D observations, (b) local navigation in random obstacle environment, (c) trajectory optimization for visual tracking. The experiments were performed on a laptop with a 2.5GHz Intel Core i5 CPU. The code and demo video is available at:

https://github.com/WilliamLiPro/TrajectoryEstimation.

### A. Typical Trajectory Estimation On 3D Observations

To verify the convergence and precision of AdaTE, we test it on 3 standard 3D trajectories with different characteristics, where the first trajectory is a cruising route with minimal maneuvers; the second one is a swaying route with frequently varying speed; the third one is a snake-like route with multiple maneuvers and lost observations (from time 250 to 300). For each standard trajectory, there are two kinds of observations: one produced with Gaussian noise and the other with additional partial drift. In all 6 groups of experiments, all tested algorithms share the same and biased prior statistics.

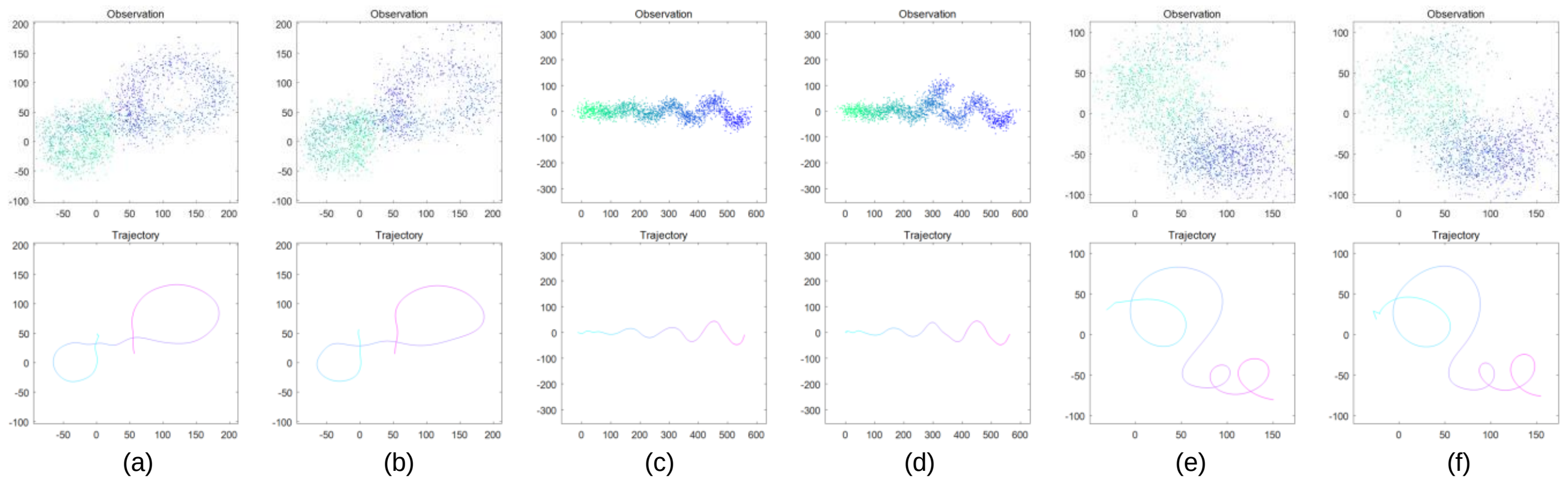

Fig. 3. The observations and results of AdaTE in 3D trajectory estimation task. The figures on the first row show the observations under different conditions, while the second row presents the results corresponding to the observations above. The coloring from red to blue is consistent with the time order. Groups (a) and (b) are based on cruising route, groups (c) and (d) are based on a swaying route, and groups (e) and (f) are based on a snake-like route with missing observations. Observation drift is added to (b), (d) and (f).

We choose EKF, UKF and sparse MAP as contrast algorithms. For the EKF and UKF, there are two different results with constant acceleration model (CA) and PLS model. For sparse MAP, the applied dynamic model is constant turning model (CT). Estimation error is measured with the normalized RMSE defined as $\frac{\text{RMSE}_{\text{result}}}{\text{RMSE}_{\text{observation}}}$.

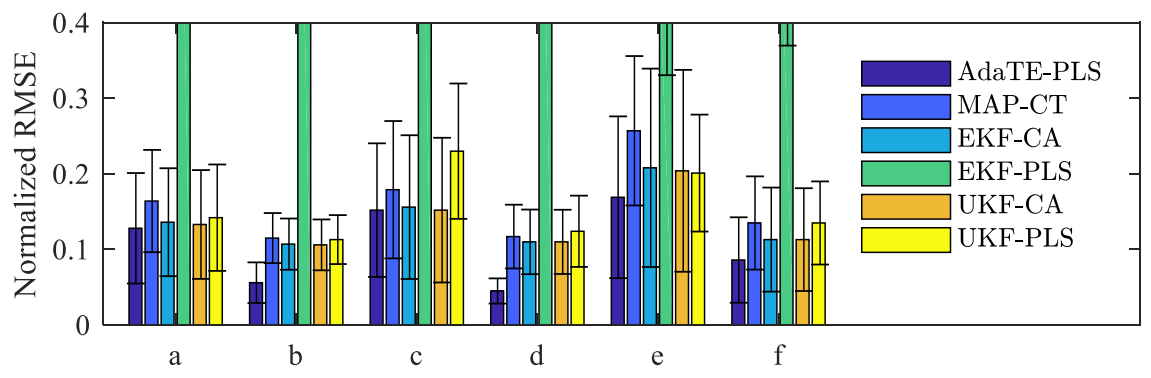

Fig. 4. The normalized RMSE of algorithms in each group of trajectory estimation, where (a)-(f) is the order of group. The bars represent the mean normalized RMSE of each algorithm, while the error lines represent the range decided by standard deviations. Note that the EKF-PLS has a divergent result in each group, and the normalized RMSE is beyond the range of the table.

TABLE I
STATISTICS OF THE RMSE IN THE TRAJECTORY ESTIMATION TASK

| Methods | Cruise | | Swaying | | Snake-like | |
|---|---|---|---|---|---|---|
| Observation drift | no | yes | no | yes | no | yes |
| MAP-CT | 0.164 | 0.115 | 0.179 | 0.117 | 0.257 | 0.135 |
| EKF-CA | 0.136 | 0.107 | 0.156 | 0.110 | 0.208 | 0.113 |
| EKF-PLS | Div | Div | Div | Div | Div | Div |
| UKF-CA | 0.133 | 0.106 | 0.152 | 0.110 | 0.204 | 0.113 |
| UKF-PLS | 0.142 | 0.113 | 0.230 | 0.124 | 0.201 | 0.135 |
| AdaTE-PLS | **0.128** | **0.056** | **0.152** | **0.045** | **0.169** | **0.086** |
| Observation RMSE | 24.78 | 58.89 | 23.65 | 91.25 | 14.19 | 28.02 |

The error used to evaluate the algorithms is the mean normalized RMSE. The RMSE of observation is given as the absolute value while RMSE of each algorithm is the normalized value by that of observation. The bold number in each group is the smallest mean normalized RMSE, and the "Div" denotes that the normalized RMSE is invalid because of the divergence. AdaTE-PLS obtains the minimum error in all groups.

The observations and the corresponding results of AdaTE are shown in Fig. 3, where the figures on the first row show the observations under different conditions, and the figures on the second row are the estimated trajectory corresponding to the observations above. Parts of the observations are drifted in Fig. 3 (b), (d) and (f). It should be noted that the AdaTE produces smooth and accurate estimation in every listed situations, even both of the missing observation and observation drift occurs in group (f).

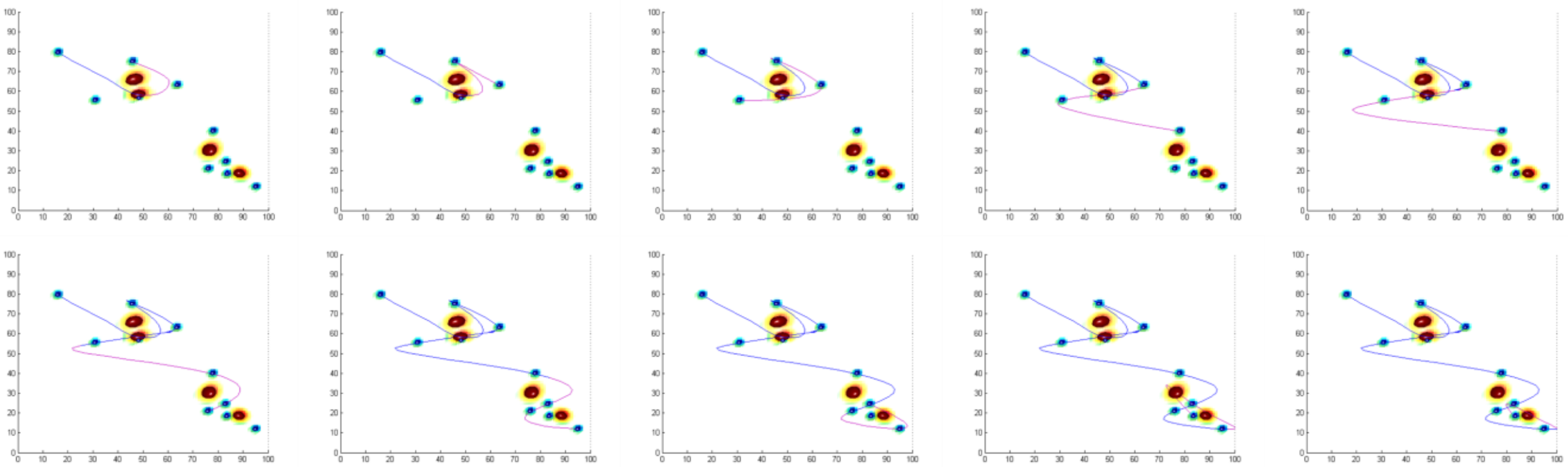

Fig. 5. Overview of the 3D navigation at 10 representative moments, where path planning is adjusted online according to the position of next nodes and real time obstacles. The images on the top line from left to right are overviews at time 10, 11, 21, 31, and 34. The bottom images are overviews at time 41, 51, 61, 71, and 73. The dark blue curve is the historical path, and the purple curve is the local navigation plan. The dark blue ellipsoids are the covariance of key points; the sky blue region around the key points is the nearby region within two standard derivations. The dark red ellipsoids are the restricted area of the obstacles, while the yellow regions around are the danger areas.

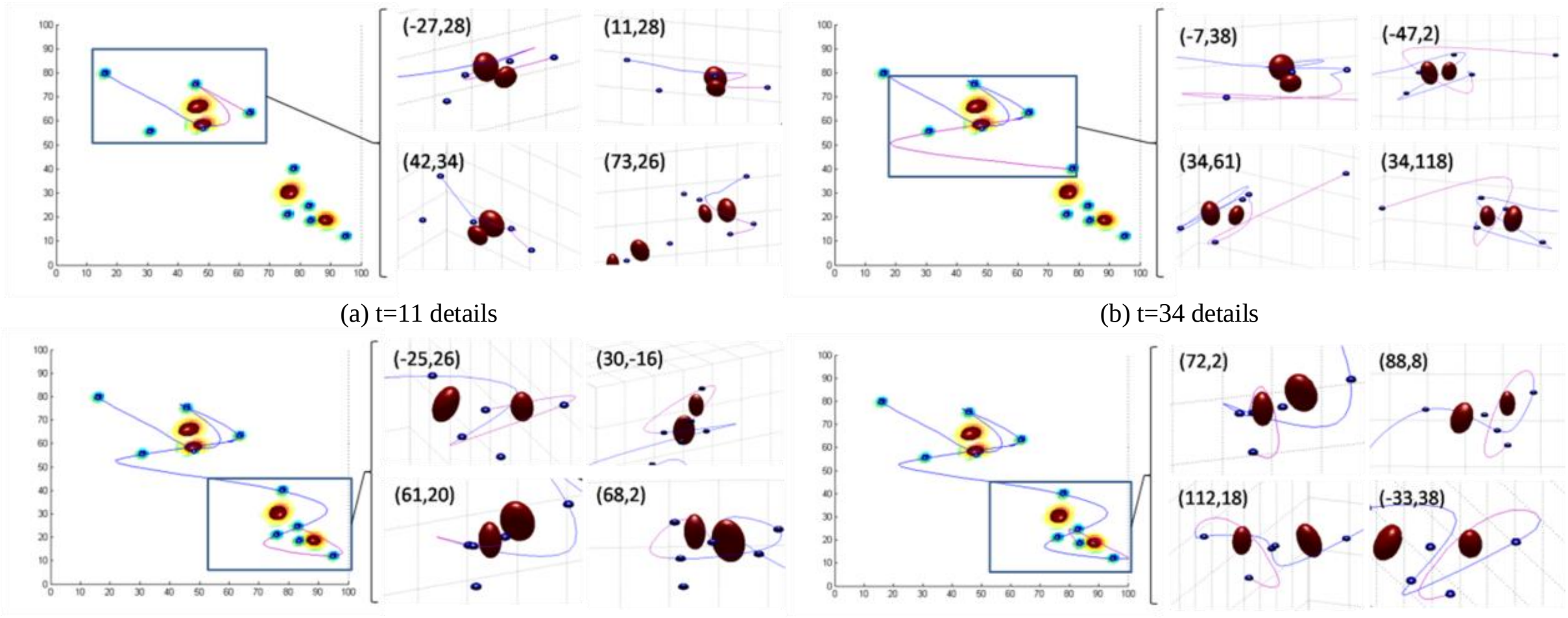

(a) t=11 details

(b) t=34 details

(c) t=61 details

(d) t=73 details

Fig. 6. The regional details of obstacle avoidance at 4 representative moments, where subgraphs (a)-(d) correspond to time 11, 34, 61 and 73. Each subgraph shows part of details from 4 different views, where the corresponding viewing angle is indicated at the top left. The dark blue curve is the historical path, and the purple curve is the planned path.

### B. Local Navigation In Random Obstacle Environment

The topological path planning is commonly applied in large scale navigation [40][41]. After getting the node orders, the best path between any two nodes should be rapidly found out for local navigation. In an environment with obstacles, for the safety and flexibility, the navigation should be online adjusted according to real-time situation. For the stability, the object ought to have fewer maneuvers while avoiding obstacles. This is a suitable task for AdaTE.

In this task, the object is assigned to avoid the random obstacles and to online update the smooth trajectory to the next two nodes, thus object can safely and punctually arrive at the planned nodes with minimal maneuverability.

Setting $\boldsymbol{P}$ be the set of key point nodes, $\boldsymbol{\Omega}$ be the set of obstacles, and $t_k$ be the time of arrival at upcoming node, the MAP of the plan can be described as

$$\boldsymbol{S}_k^{(k:t_k)} = \underset{\boldsymbol{X}_k^{(k:t_k)}}{\arg\max}\, \mathcal{L}(\boldsymbol{S}_k^{(k:t_k)}; \boldsymbol{S}_{k-1}^{(k:t_{k-1})}, \boldsymbol{P}, \boldsymbol{\Omega}) \tag{42}$$

where

$$\begin{aligned} &\mathcal{L}(\boldsymbol{S}_k^{(k:t_k)}; \boldsymbol{S}_{k-1}^{(k:t_{k-1})}, \boldsymbol{P}, \boldsymbol{\Omega}) = \\ &\log P(\boldsymbol{S}_k^{(k:t_k)} \mid \boldsymbol{S}_{k-1}^{(k:t_{k-1})}) + \log P(\boldsymbol{P} \mid \boldsymbol{S}_k^{(k:t_k)}) + \\ &\log P_a(\boldsymbol{\Omega} \mid \boldsymbol{S}_k^{(k:t_k)}) \end{aligned} \tag{43}$$

and

$$P_a(\boldsymbol{\Omega} \mid \boldsymbol{S}_k^{(k:t_k)}) = \prod_{i=k}^{t_k} \left(1 - P_h(\boldsymbol{\Omega} \mid \boldsymbol{S}_k^{(i)})\right) \tag{44}$$

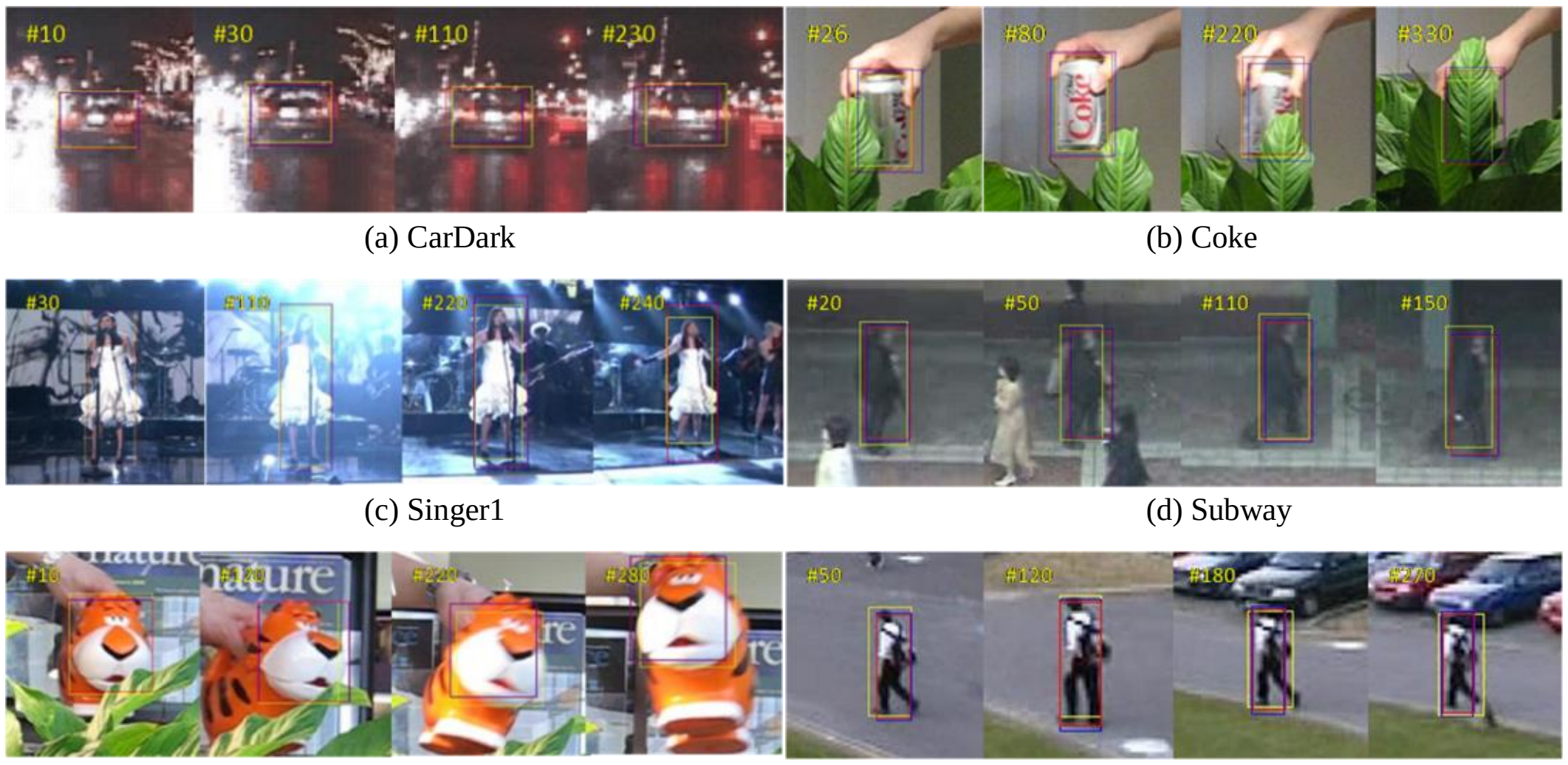

(a) CarDark (b) Coke

(c) Singer1 (d) Subway

(e) Tiger1 (f) Walking

Fig. 7. Some screenshots in visual tracking experiments, including the result of original fDSST (blue box), the result estimated by fDSST-AdaTE (red box), and the ground truth (yellow box). Note that in most of the images, fDSST-AdaTE obtaines a slightly better result than fDSST.

TABLE II
MEAN OVERLAP ON EIGHT SCENES FROM VOT 2013 DATASET

| | fDSST | fDSST-AdaTE | Increase |
|---|---|---|---|
| *vot-cardark* | 0.7422 | **0.7507** | 1.16% |
| *vot-coke* | 0.5589 | **0.5604** | 0.27% |
| *vot-tiger1* | 0.7337 | **0.7349** | 0.17% |
| *vot-tiger2* | **0.6759** | 0.6723 | -0.53% |
| *vot-walking* | 0.7257 | **0.7333** | 1.05% |
| *vot-subway* | 0.6880 | **0.6914** | 0.50% |
| *vot-singer1* | 0.8042 | **0.8149** | 1.33% |
| *vot-dudek* | 0.8034 | **0.8062** | 0.35% |

A comparison of fDSST and fDSST-AdaTE in visual tracking task on 8 scenes from the VOT 2013 dataset. The index for algorithm evaluation is mean overlap. The table shows that in most scenes, fDSST-AdaTE has a better result compared with fDSST.

where $P(\boldsymbol{P} \mid \boldsymbol{S}_k^{(k:t_k)})$ is the probability of arriving at planed nodes, $P_a(\boldsymbol{\Omega} \mid \boldsymbol{S}_k^{(k:t_k)})$ is the probability of successful obstacles avoidance during the time $(k, t_k)$, and $P_h(\boldsymbol{\Omega} \mid \boldsymbol{S}_k^{(i)})$ is the unblocked probability at moment $i$.

Noticing that the likelihood defined in (43) is an extended form of (28) with an additional unblocked item, the corresponding real-time navigation could be solved by extended AdaTE. The overview of 10 representative moments during online navigation is shown in Fig. 5. We can observe that when the object arriving at next node, a new node in plan is taken into consideration and the original path planning is slightly adjusted (compare the pink curve at time 10 with that at time 11). Another detail shows that the path planning at time 34 is adjusted compared with the planning at time 31. The reason is that the original path fails to avoid the obstacle; therefore, at time 34 it was fixed to avoid from the bottom of obstacle (for details, please turn to Fig. 6 (b)). The evasive maneuvers extend the path and increasing the planning speed, leading to a larger turning radius. A similar situation can be found between time 71 and 73.

The overview of the 3D path cannot display the details of the obstacle avoidance, especially when the object passes obstacle from above or below, the object appearing to be intersecting them. To prevent from misunderstanding, several details of the obstacle avoidance are presented in Fig. 6. In general, AdaTE uses minimum maneuverability under the constraint of the dynamic model to avoid the obstacles while successfully arriving at the planed nodes. In this experiment, the navigation converges in 3 loops of iterations for a single nearby obstacle. And it needs no more than 4 iterations for two obstacles.

In environment with sparsely distributed obstacles, AdaTE has an excellent real time performance by an $O(n)$ computational complexity, where $n$ is the number of nearby obstacles. As a comparison, some powerful obstacle avoidance approaches such as Cooperative Collision Avoidance (CCA) [42] has obvious higher complexity $O(2n^2)$. The reason comes from their motivations: AdaTE is an online iterative algorithm, thus the result is adjusted based on previous plan, while CCA is an offline algorithm without online iteration

### *C. Trajectory Optimization For Visual Tracking*

To test the application of AdaTE in visual tracking, we choose 8 scenes in VOT 2013 with illumination variations (CarDark and Singer1), scale variations (Singer1) and partial blocking (Coke and Tiger1) for the experiment. The fDSST [43] is applied as the basic tracking algorithm, and the mean overlap is chosen as the evaluation index. Parts of the screenshots are presented in Fig. 7. It is obvious that in most scenes, the AdaTE achieves a slight improvement in accuracy compared with basic fDSST.

The mean overlap of fDSST and fDSST-AdaTE is shown in TABLE II. In most scenes, fDSST-AdaTE is superior to the basic fDSST with slight advantage. We find that as a model-based algorithm, AdaTE is robust to the shaky observations compared with the original tracking algorithm. However, AdaTE cannot distinguish small shakiness from noise with the shakiness in real motion, thus leading to the extra estimation error in certain scenes, such as Tiger2. Compared with applying to a smooth tracking algorithm like DSST, AdaTE is likely to achieve more obvious improvement for a wobbly tracking algorithm such as SAMF [3].

# VI. Conclusion

This paper investigates the online trajectory estimation of non-cooperative or maneuvering under the condition of regional correlations, biased prior statistics and observations drift. And the works focus on two aspects: the dynamic model and robust estimation algorithm.

To estimate the trajectory without the dedicated dynamic model, we learn from the previous works and propose the power-limited steering model, which is a natural combination of instantaneous power theory and instantaneous angular velocity theory from Newtonian mechanics. The contribution has two aspects: a) overcoming the infinite speed problem and convergent in prediction; B) sustaining few parameters to describe the switch of typical states.

To achieve robust and efficient trajectory estimation in challenging situations such as the biased prior statistics and the observations drift, we propose the adaptive trajectory estimation algorithm. The contribution of AdaTE is threefold a) the adaptation of transition covariance matrices improves the estimation robustness to the model error and prior statistics; b) the correction of observation covariance matrices improves the estimation robustness to the observation drift; c) the truncation time ensures the calculation is restricted to the necessary part, which avoids the linear improvement of calculation.

To evaluate the performance of AdaTE-PLS in different tasks, we conduct three experiments: a) typical trajectory estimation on 3D observations, b) local navigation over node planning, c) trajectory optimization for visual tracking. By testing on three standard 3D routes with different characteristics, the convergence and precision of AdaTE is verified. Experiment b) and c) demonstrate that with slight modification, AdaTE can be widely applied in tasks such as local navigation under random obstacls, and trajectory optimization in visual tracking.

In the future, we will explore better dynamic model which is simpler, less calculation and more precise for maneuvering object. Furthermore, we hope to extend the AdaTE algorithm to a wider probability distribution, such as the exponential family, to develop a more significant and general regional optimization method.

# Appendix

## A. Solving The Predicted Velocity From PLS Model When Ignoring The Perturbation

When ignoring the perturbation of $p$ in differential equation (15) during a short time $(t, t+\tau)$, the axial velocity satisfies:

$$\frac{\|\boldsymbol{v}\|}{\alpha\|\boldsymbol{v}\|^2+\beta\|\boldsymbol{v}\|-p} d\|\boldsymbol{v}\| = -dt \tag{45}$$

where $p$ has a fixed value during $(t, t+\tau)$.

Rewrite (45) as:

$$\left(\frac{\varsigma_1}{\|\boldsymbol{v}\|+\varsigma_1}-\frac{\varsigma_2}{\|\boldsymbol{v}\|+\varsigma_2}\right) d\|\boldsymbol{v}\| = -\gamma dt \tag{46}$$

where

$$\begin{cases} \varsigma_1 = \dfrac{\beta+\sqrt{\beta^2+4\alpha p^*}}{2\alpha} \\ \varsigma_2 = \dfrac{\beta-\sqrt{\beta^2+4\alpha p^*}}{2\alpha} \end{cases} \tag{47}$$

$$\gamma = \sqrt{\beta^2+4\alpha p} \tag{48}$$

When ignoring the perturbation of $p$ during time $(t, t+\tau)$. The biased estimation of $\|\boldsymbol{v}\|$ can be solved from (46):

$$\left[\varsigma_1 \log\left(\|\boldsymbol{v}\|+\varsigma_1\right)-\varsigma_2 \log\left(\|\boldsymbol{v}\|+\varsigma_2\right)\right]\Big|_t^{t+\tau} = -\gamma\tau \tag{49}$$

Calculate the exponential on both side of (49), we get,

$$\left(\frac{\|\hat{\boldsymbol{v}}(t+\tau)\|+\varsigma_1}{\|\boldsymbol{v}(t)\|+\varsigma_1}\right)^{\varsigma_1}\left|\frac{\|\hat{\boldsymbol{v}}(t+\tau)\|+\varsigma_2}{\|\boldsymbol{v}(t)\|+\varsigma_2}\right|^{-\varsigma_2}=e^{-\gamma\tau} \tag{50}$$

Then consider the orientation of velocity. From (50) the orientation of velocity satisfies,

$$\frac{d}{dt}\frac{\boldsymbol{v}}{\|\boldsymbol{v}\|}=\left[\boldsymbol{c}_{\mathrm{T}}\right]_{\times}\frac{\boldsymbol{v}}{\|\boldsymbol{v}\|} \tag{51}$$

Solving (51), we can get the orientation of velocity,

$$\frac{\hat{\boldsymbol{v}}(t+\tau)}{\|\hat{\boldsymbol{v}}(t+\tau)\|}=\exp\left(\tau\left[\boldsymbol{c}_{\mathrm{T}}(t)\right]_{\times}\right)\frac{\boldsymbol{v}(t)}{\|\boldsymbol{v}(t)\|} \tag{52}$$

Combining (50) and (52), the relationship between previous velocity $\boldsymbol{v}(t)$ and predicted velocity $\hat{\boldsymbol{v}}(t+\tau)$ satisfies,

$$\begin{cases}\left(\frac{\|\hat{\boldsymbol{v}}(t+\tau)\|+\varsigma_1}{\|\boldsymbol{v}(t)\|+\varsigma_1}\right)^{\varsigma_1}\left|\frac{\|\hat{\boldsymbol{v}}(t+\tau)\|+\varsigma_2}{\|\boldsymbol{v}(t)\|+\varsigma_2}\right|^{-\varsigma_2}=e^{-\gamma\tau} \\ \frac{\hat{\boldsymbol{v}}(t+\tau)}{\|\hat{\boldsymbol{v}}(t+\tau)\|}=\exp\left(\tau\left[\boldsymbol{c}_{\mathrm{T}}(t)\right]_{\times}\right)\frac{\boldsymbol{v}(t)}{\|\boldsymbol{v}(t)\|}\end{cases} \tag{53}$$

### *B. The Compensation Of Velocity Bassed On The Renormalization Group*

Setting the axial speed $\upsilon=\|\boldsymbol{v}\|$. From (15), the function of axial acceleration is,

$$a_{A0}\left(\upsilon\right)=\frac{p}{\upsilon}-\alpha\upsilon-\beta \tag{54}$$

At reference time $t$, the zero order axial speed is $\upsilon_0=\|\boldsymbol{v}(t)\|$, from (54) the first order renormalization of $\upsilon$ satisfies,

$$\begin{aligned}\upsilon_1(t+\tau)&=\upsilon_0+\int_t^{t+\tau}a_A\left(\upsilon_0\right)ds \\ &=\upsilon_0+\frac{\tau p(t)+r^{(1)}(\tau)}{\upsilon_0}-\tau\left(\alpha\upsilon_0+\beta\right)\end{aligned} \tag{55}$$

where $r^{(1)}(\tau)=\int_0^{\tau}\int_0^{s}u_A(\iota)dsd\iota$ obeys $\mathcal{N}\left\{0,\frac{1}{3}\tau^3D_A\right\}$.

Put $\upsilon_1$ into (54) we can get $a_{A0}\left(\upsilon_1\right)$. Then plug $a_{A0}\left(\upsilon_1\right)$ into (55), we will obtain the second order renormalization of $\upsilon$,

$$\begin{aligned}\upsilon_2(t+\tau)&=\upsilon_0+\int_t^{t+\tau}a_A\left(\upsilon_1\right)ds \\ &=\upsilon_0+\int_0^{\tau}\frac{p(t)+r(s)}{\upsilon_0+\frac{sp(t)+r^{(1)}(s)}{\upsilon_0}-s\left(\alpha\upsilon_0+\beta\right)}ds+\Omega(t+\tau)\end{aligned} \tag{56}$$

where $r(s)=\int_0^{s}u_A(\iota)d\iota$ obeys $\mathcal{N}\left\{0,\tau D_A\right\}$

Noticing that when $\tau$ is a small number, and $\upsilon_0\neq 0\vee p(t)\neq 0$, is a small number, (56) can be rewritten as,

$$\begin{aligned}\upsilon_2(t+\tau)&=\int_0^{\tau}\frac{\upsilon_0\left(p(t)+r(s)\right)\left[\upsilon_0^2+sp(t)-s\upsilon_0\left(\alpha\upsilon_0+\beta\right)-r^{(1)}(s)\right]}{\left(\upsilon_0^2+sp(t)-s\upsilon_0\left(\alpha\upsilon_0+\beta\right)\right)^2-\left[r^{(1)}(s)\right]^2}ds+\Omega(t+\tau) \\ &\approx-\int_0^{\tau}\frac{\upsilon_0 r(s)r^{(1)}(s)}{\left(\upsilon_0^2+sp(t)-s\upsilon_0\left(\alpha\upsilon_0+\beta\right)\right)^2}ds+\Omega(t+\tau)\end{aligned} \tag{57}$$

where $\Omega(t+\tau)$ is the part of velocity ignoring disturbance, which is equal to $\|\hat{\boldsymbol{v}}(t+\tau)\|$ in this paper.

Based on (57), the expectation of $\upsilon_2(t+\tau)$ satisfies,

$$\begin{aligned}\mathrm{E}[\upsilon_2(t+\tau)] &\approx -\upsilon_0 \int_0^{\tau} \frac{\mathrm{E}[r(s)r^{(1)}(s)]}{\left(\upsilon_0^2 + sp(t) - s\upsilon_0\left(\alpha\upsilon_0 + \beta\right)\right)^2} ds + \Omega(t+\tau) \\ &= \Delta\upsilon(t,t+\tau) + \Omega(t+\tau)\end{aligned} \tag{58}$$

where

$$\mathrm{E}[r(s)r^{(1)}(s)] = \mathrm{E}\left[\int_0^s\int_0^{\rho} u_A(\iota)d\iota d\rho \bullet \int_0^s u_A(\iota)d\iota\right] = \frac{1}{2}s^2 D_A \tag{59}$$

and $\Delta\upsilon(t,t+\tau)$ is the compensation part of $\mathrm{E}[\upsilon(t+\tau)]$,

$$\begin{aligned}&\Delta\upsilon(t,t+\tau) = \\ &-\frac{\upsilon_0 D_A}{2\mu(t)}\left[\tau - \frac{2\upsilon_0^2}{\mu(t)}\log\frac{\upsilon_0^2 + \tau\mu(t)}{\upsilon_0^2} + \frac{\upsilon_0^2\tau}{\upsilon_0^2 + \tau\mu(t)}\right]\end{aligned} \tag{60}$$

where $\mu(t) = p(t) - \upsilon_0\left(\alpha\upsilon_0 + \beta\right)$.

From (16), we know the angular velocity $\frac{d}{dt}\frac{\boldsymbol{v}}{\|\boldsymbol{v}\|}$ is linear with $\frac{\boldsymbol{v}}{\|\boldsymbol{v}\|}$ and $\left[\boldsymbol{c}_T\right]_\times$, so the $\boldsymbol{u}_T$ will not act on the expectation of orientation, orientation doesn't need any compensation.

### *C. The Approximate Calculation Of Transition Matrix*

To match the requirement of the optimization algorithm in next section, the state transition function should be written in a linear form:

$$\boldsymbol{s}_{k+1} = \boldsymbol{\Phi}_{k,k+1}\boldsymbol{s}_k + \boldsymbol{\varepsilon}_{k,k+1} \tag{61}$$

where $\boldsymbol{s} = \left[\boldsymbol{x}^{\mathrm{T}}\ \boldsymbol{v}^{\mathrm{T}}\ p\ \boldsymbol{c}_T^{\mathrm{T}}\right]^{\mathrm{T}}$, $\boldsymbol{s}_k$ and $\boldsymbol{s}_{k+1}$ is the state vector at time $k$ and $k+1$, $\boldsymbol{\varepsilon}_{k,k+1}$ is the transition error.

To work out the $\boldsymbol{\Phi}_{k,k+1}$ and $\boldsymbol{\varepsilon}_{k,k+1}$ from velocity evolution, the continuous time form of (61) should be solved at first,

$$\boldsymbol{s}(t+\tau) = \boldsymbol{\Phi}(t,t+\tau)\boldsymbol{s}(t) + \boldsymbol{\varepsilon}(t,t+\tau) \tag{62}$$

where

$$\begin{aligned}&\boldsymbol{\Phi}(t,t+\tau) = \\ &\begin{bmatrix} \boldsymbol{I} & \boldsymbol{\Phi}_{12}(t,t+\tau) & \boldsymbol{\Phi}_{13}(t,t+\tau) & \boldsymbol{\Phi}_{14}(t,t+\tau) \\ \boldsymbol{0} & \boldsymbol{\Phi}_{22}(t,t+\tau) & \boldsymbol{\Phi}_{23}(t,t+\tau) & \boldsymbol{\Phi}_{24}(t,t+\tau) \\ \boldsymbol{0} & \boldsymbol{0} & 1 & \boldsymbol{0} \\ \boldsymbol{0} & \boldsymbol{0} & \boldsymbol{0} & \boldsymbol{I} \end{bmatrix}\end{aligned} \tag{63}$$

According to (17)-(20), the velocity at time $t+\tau$ can be mentioned as,

$$\begin{aligned}\boldsymbol{v}(t+\tau) &= \bar{\boldsymbol{v}}(t+\tau) + \delta\boldsymbol{v}(t+\tau) \\ &= \exp\left(\tau\left[\boldsymbol{c}_{\mathrm{T}}(t)\right]_\times\right)\frac{|\bar{\boldsymbol{v}}(t+\tau)|}{|\boldsymbol{v}(t)|}\boldsymbol{v}(t) + \delta\boldsymbol{v}(t,t+\tau)\end{aligned} \tag{64}$$

where $\delta\boldsymbol{v}(t,t+\tau)$ is the error of velocity transition.

Combining (17) and (19), the axial velocity can be decomposed as,

$$|\bar{\boldsymbol{v}}(t+\tau)| = \left(|\boldsymbol{v}(t)| + \varsigma_1\right)e^{-\lambda\tau}g_{(t,t+\tau)} - \varsigma_1 + \Delta\upsilon(t,t+\tau) \tag{65}$$

where

$$\begin{cases} g_{(t,t+\tau)} = \left|\dfrac{|\hat{\boldsymbol{v}}(t+\tau)| + \varsigma_2}{|\boldsymbol{v}(t)| + \varsigma_2}\right|^{\frac{\varsigma_2}{\varsigma_1}} \\ \lambda = \dfrac{\gamma}{\varsigma_1} \end{cases} \tag{66}$$

Plug (65) into (64), the $\boldsymbol{v}(t+\tau)$ satisfies,

$$\boldsymbol{v}(t+\tau) = \frac{\left(|\boldsymbol{v}(t)| + \varsigma_1\right)e^{-\lambda\tau}g_{(t,t+\tau)} - \varsigma_1 + \Delta\upsilon(t,t+\tau)}{|\boldsymbol{v}(t)|}\exp\left(\tau\left[\boldsymbol{c}_{\mathrm{T}}(t)\right]_\times\right)\boldsymbol{v}(t) + \delta\boldsymbol{v}(t+\tau) \tag{67}$$

To derive the transition matrix $\boldsymbol{\Phi}(t,t+\tau)$ in (63), we should expand the velocity evolution (67) into a linear function of $\boldsymbol{v}(t)$, $p(t)$ and $\boldsymbol{c}_T(t)$,

$$\boldsymbol{v}(t+\tau) \approx \left( e^{-\lambda\tau} g_{(t,t+\tau)} + \frac{\Delta \upsilon(t,t+\tau)}{|\boldsymbol{v}(t)|} \right) \boldsymbol{v}(t) + h_{(t,t+\tau)} \left(1 - e^{-\lambda\tau} g_{(t,t+\tau)}\right) \frac{\boldsymbol{v}(t)}{|\boldsymbol{v}(t)|} p(t) \\ -\tau \varphi_{(t,t+\tau)} [\boldsymbol{v}(t)]_{\times} \boldsymbol{c}_T(t) + \delta \boldsymbol{v}(t+\tau) \tag{68}$$

where

$$\begin{cases} h_{(t,t+\tau)} = \dfrac{2}{\beta - \gamma} \\ \varphi_{(t,t+\tau)} = e^{-\lambda\tau} g(t,t+\tau) + \dfrac{\Delta \upsilon(t,t+\tau)}{|\boldsymbol{v}(t)|} + \dfrac{h_{(t,t+\tau)} p(t)}{|\boldsymbol{v}(t)|} \left(1 - e^{-\lambda\tau} g_{(t,t+\tau)}\right) \end{cases} \tag{69}$$

From (68), we can obtain the submatrix of $\boldsymbol{\Phi}(t,t+\tau)$ about velocity in (63):

$$\begin{cases} \boldsymbol{\Phi}_{22}(t,t+\tau) \cong \left( e^{-\lambda\tau} g_{(t,t+\tau)} + \dfrac{\Delta \upsilon(t,t+\tau)}{|\boldsymbol{v}(t)|} \right) \boldsymbol{I} \\ \boldsymbol{\Phi}_{23}(t,t+\tau) \cong \left(1 - e^{-\lambda\tau} g_{(t,t+\tau)}\right) \dfrac{h_{(t,t+\tau)} \boldsymbol{v}^*(t)}{|\boldsymbol{v}^*(t)|} \\ \boldsymbol{\Phi}_{24}(t,t+\tau) \cong -\tau \varphi_{(t,t+\tau)} [\boldsymbol{v}^*(t)]_{\times} \end{cases} \tag{70}$$

Then from $\boldsymbol{x}(t+\tau) = \boldsymbol{x}(t) + \int_{s=t}^{t+\tau} \boldsymbol{v}(t) dt$, the submatrix of $\boldsymbol{\Phi}(t,t+\tau)$ about position satisfies,

$$\begin{cases} \boldsymbol{\Phi}_{12}(t,t+\tau) = \tau \left( 0.5 + 0.5 e^{-\lambda\tau} g_{(t,t+\tau)} + \dfrac{\Delta \upsilon(t,t+\tau)}{2|\boldsymbol{v}(t)|} \right) \boldsymbol{I} \\ \boldsymbol{\Phi}_{13}(t,t+\tau) = \tau \left(0.5 - 0.5 e^{-\lambda\tau} g_{(t,t+\tau)}\right) \dfrac{h_{(t,t+\tau)} \boldsymbol{v}^*(t)}{|\boldsymbol{v}^*(t)|} \\ \boldsymbol{\Phi}_{14}(t,t+\tau) = -\dfrac{1}{2} \tau^2 \varphi_{(t,t+\tau)} [\boldsymbol{v}^*(t)]_{\times} \end{cases} \tag{71}$$

### *D. The Approximate Calculation Of Transition Covariance*

Because the expectation of velocity perturbation have been compensated in (19), $\boldsymbol{\varepsilon}(t,t+\tau)$ will be a zero-mean random vector. Thus $\boldsymbol{\varepsilon}(t,t+\tau) \sim \mathcal{N}\left\{\boldsymbol{0}, \boldsymbol{Q}(t,t+\tau)\right\}$,

$$\boldsymbol{Q}(t,t+\tau) = Cov[\boldsymbol{\varepsilon}(t,t+\tau)] = \int_{t}^{t+\tau} \boldsymbol{\Phi}(t,s) \boldsymbol{D}_u \boldsymbol{\Phi}^{\mathrm{T}}(t,s) ds \tag{72}$$

where $\boldsymbol{D}_u$ is the power spectral density of $\boldsymbol{u}$, which satisfies,

$$\boldsymbol{D}_u = \int_{t}^{t+\tau} \mathrm{E}[\boldsymbol{u}(s_1) \boldsymbol{u}^{\mathrm{T}}(s_2)] ds_1 = diag\{0 \quad 0 \quad D_A \quad \boldsymbol{D}_T\} \tag{73}$$

Substituting (63) into (72), the transition covariance will be,

$$\boldsymbol{Q}(t,t+\tau) = \\ \begin{bmatrix} \boldsymbol{\Sigma}_{11}(t,t+\tau) & \boldsymbol{\Sigma}_{12}(t,t+\tau) & \boldsymbol{\Sigma}_{13}(t,t+\tau) & \boldsymbol{\Sigma}_{14}(t,t+\tau) \\ \boldsymbol{\Sigma}_{12}^T(t,t+\tau) & \boldsymbol{\Sigma}_{22}(t,t+\tau) & \boldsymbol{\Sigma}_{23}(t,t+\tau) & \boldsymbol{\Sigma}_{24}(t,t+\tau) \\ \boldsymbol{\Sigma}_{13}^T(t,t+\tau) & \boldsymbol{\Sigma}_{23}^T(t,t+\tau) & \boldsymbol{\Sigma}_{33}(t,t+\tau) & \boldsymbol{0} \\ \boldsymbol{\Sigma}_{14}^T(t,t+\tau) & \boldsymbol{\Sigma}_{24}^T(t,t+\tau) & \boldsymbol{0} & \boldsymbol{\Sigma}_{44}(t,t+\tau) \end{bmatrix} \tag{74}$$

where

$$
\begin{cases}
\boldsymbol{\Sigma}_{11}(t,t+\tau)=\int_t^{t+\tau} D_A\boldsymbol{\Phi}_{13}(t,s)\boldsymbol{\Phi}_{13}^T(t,s)+\boldsymbol{\Phi}_{14}(t,s)\boldsymbol{D}_T\boldsymbol{\Phi}_{14}^T(t,s)ds \\
\boldsymbol{\Sigma}_{12}(t,t+\tau)=\int_t^{t+\tau} D_A\boldsymbol{\Phi}_{13}(t,s)\boldsymbol{\Phi}_{23}^T(t,s)+\boldsymbol{\Phi}_{14}(t,s)\boldsymbol{D}_T\boldsymbol{\Phi}_{24}^T(t,s)ds \\
\boldsymbol{\Sigma}_{13}(t,t+\tau)=\int_t^{t+\tau} D_A\boldsymbol{\Phi}_{13}(t,s)ds \\
\boldsymbol{\Sigma}_{14}(t,t+\tau)=\int_t^{t+\tau} \boldsymbol{\Phi}_{14}(t,s)\boldsymbol{D}_T ds
\end{cases}
\tag{75}
$$

and

$$
\begin{cases}
\boldsymbol{\Sigma}_{22}(t,t+\tau)=\int_t^{t+\tau} D_A\boldsymbol{\Phi}_{23}(t,s)\boldsymbol{\Phi}_{23}^T(t,s)+\boldsymbol{\Phi}_{24}(t,s)\boldsymbol{D}_T\boldsymbol{\Phi}_{24}^T(t,s)ds \\
\boldsymbol{\Sigma}_{23}(t,t+\tau)=\int_t^{t+\tau} D_A\boldsymbol{\Phi}_{23}(t,s)ds \\
\boldsymbol{\Sigma}_{24}(t,t+\tau)=\int_t^{t+\tau} \boldsymbol{\Phi}_{24}(t,s)\boldsymbol{D}_T ds
\end{cases}
\tag{76}
$$

and

$$
\begin{cases}
\boldsymbol{\Sigma}_{33}(t,t+\tau)=\tau D_A \\
\boldsymbol{\Sigma}_{44}(t,t+\tau)=\tau \boldsymbol{D}_T
\end{cases}
\tag{77}
$$

After ignoring the high order small quantity of $\tau$, the approximation of each submatrix is,

$$
\begin{cases}
\boldsymbol{\Sigma}_{11}(t,t+\tau)\approx\frac{\tau^3}{12}D_A\left(1-e^{-\lambda\tau}g_{(t,t+\tau)}\right)^2\frac{h_{(t,t+\tau)}^2}{|\boldsymbol{v}(t)|^2}\boldsymbol{v}(t)\boldsymbol{v}(t)^{\mathrm{T}}-\frac{1}{20}\tau^5\varphi_{(t,t+\tau)}^2[\boldsymbol{v}(t)]_\times\boldsymbol{D}_T[\boldsymbol{v}(t)] \\
\boldsymbol{\Sigma}_{12}(t,t+\tau)\approx\frac{\tau^2}{4}D_A(1-e^{-\lambda\tau}g_{(t,t+\tau)})^2\frac{h_{(t,t+\tau)}^2}{|\boldsymbol{v}(t)|^2}\boldsymbol{v}(t)\boldsymbol{v}(t)^{\mathrm{T}}-\frac{1}{8}\tau^4\varphi_{(t,t+\tau)}^2[\boldsymbol{v}(t)]_\times\boldsymbol{D}_T[\boldsymbol{v}(t)] \\
\boldsymbol{\Sigma}_{13}(t,t+\tau)=\frac{\tau^2}{4}D_Ag_{(t,t+\tau)}\frac{h_{(t,t+\tau)}}{|\boldsymbol{v}(t)|}\boldsymbol{v}(t) \\
\boldsymbol{\Sigma}_{14}(t,t+\tau)=-\frac{1}{6}\tau^3\varphi_{(t,t+\tau)}[\boldsymbol{v}(t)]_\times\boldsymbol{D}_T
\end{cases}
\tag{78}
$$

and

$$
\begin{cases}
\boldsymbol{\Sigma}_{22}(t,t+\tau)=\tau D_A(1-g_{(t,t+\tau)})^2\frac{h_{(t,t+\tau)}^2}{|\boldsymbol{v}(t)|^2}\boldsymbol{v}(t)\boldsymbol{v}(t)^{\mathrm{T}}-\frac{1}{3}\tau^3\varphi_{(t,t+\tau)}^2[\boldsymbol{v}_k]_\times\boldsymbol{D}_T[\boldsymbol{v}_k] \\
\boldsymbol{\Sigma}_{23}(t,t+\tau)=\tau D_Ag_{(t,t+\tau)}\frac{h_{(t,t+\tau)}}{|\boldsymbol{v}(t)|}\boldsymbol{v}(t) \\
\boldsymbol{\Sigma}_{24}(t,t+\tau)=-\frac{1}{2}\tau^2\varphi_{(t,t+\tau)}[\boldsymbol{v}(t)]_\times\boldsymbol{D}_T
\end{cases}
\tag{79}
$$